\documentclass[11pt]{article}

\usepackage{amsmath}
\usepackage{amssymb}
\usepackage{graphicx}

\usepackage{cite}

\usepackage[labelfont=bf,labelsep=period,justification=raggedright]{caption}
\usepackage{caption}

\makeatletter
\renewcommand{\@biblabel}[1]{\quad#1.}
\makeatother

\date{}

\begin{document}

\begin{flushleft}
{\Large \textbf{Interactions of cultures and
top people of Wikipedia \\
 from ranking of 24 language editions } }
\\ \bigskip
Young-Ho\ Eom$^{1}$,
Pablo Arag\'on$^{2}$,
David Laniado$^{2}$,
Andreas Kaltenbrunner$^{2}$,
Sebastiano Vigna$^{3}$,
Dima L.\ Shepelyansky$^{1,*}$
\\ \medskip
{1} {\it Laboratoire de Physique Th\'eorique du CNRS, IRSAMC,
Universit\'e de Toulouse, UPS, F-31062 Toulouse, France}
\\
{2} {\it Barcelona Media Foundation, Barcelona, Spain}
\\
{3} {\it Dipartimento di Informatica,
Universit\`a degli Studi di Milano, Italy}
\\
\medskip
$\ast$ Webpage: www.quantware.ups-tlse.fr/dima
\end{flushleft}

\section*{Abstract}

%
Wikipedia is a huge global repository of human knowledge,
that can be leveraged to investigate interwinements between cultures. 
With this aim, we apply methods of Markov chains and Google matrix,
for the analysis of the hyperlink networks of 24 Wikipedia language editions,
and rank all their articles by PageRank, 2DRank and CheiRank algorithms.
Using automatic extraction of people names,
we obtain the top 100 historical figures, for each edition and
for each algorithm.
We investigate their spatial, temporal, and gender
distributions in dependence of their cultural origins.
Our study demonstrates not only the existence of skewness
with local figures, mainly recognized only in their own cultures, but also
the existence of global historical figures appearing in a large number of
editions.
By determining the birth time and place of these persons,
we perform an analysis of the evolution of
such figures through 35 centuries of human history for each language,
thus recovering interactions and entanglement of cultures
over time. We also obtain the distributions of historical figures
over world countries, highlighting geographical
aspects of cross-cultural links.
Considering historical figures who appear in multiple editions as
interactions between cultures, we construct a network of cultures
and identify the most influential cultures according to this network.

\noindent

\newpage$\phantom{.}$

\section*{Introduction}


The influence of digital media on collective opinions, social
relationships, and information dynamics is growing significantly
with the advances of information technology.
On the other hand,
understanding how collective opinions are reflected
in digital media has crucial importance.
Among such a medium, Wikipedia, the open, free,
and online encyclopedia, has crucial
importance since it is not only the largest global knowledge
repository but also the biggest collaborative knowledge
platform on the Web. Thanks to its huge size, broad
coverage and ease of use, Wikipedia is currently one of
the most widely used knowledge references. However,
since its beginning, there have been
constant concerns about the reliability of Wikipedia
 because of its openness. Although professional scholars
may not be affected by a possible skewness or bias of Wikipedia, students and
the public can be affected
significantly~\cite{Rosenzweig2006,Lavsa2011}. Extensive studies
have examined the reliability of contents~\cite{Giles2005,
Rosenzweig2006,Lavsa2011}, topic coverage~\cite{Kittur2009},
vandalism~\cite{Priedhorsky2007}, and
conflict~\cite{Yasseri2012,Yasseri2013,Laniado2011} in Wikipedia.

Wikipedia is available in different language editions;  287
language editions are currently active. This indicates that
the same topic can be described in hundreds of articles
written by different language user groups.
Since language is one of the primary elements of
culture~\cite{Unesco}, collective cultural biases may be reflected
on the contents and organization of each Wikipedia edition.
Although Wikipedia adopts a ``neutral point of view"  policy
for the description of contents, aiming to provide unbiased information to the
public~\cite{NPOV}, it is natural that each language edition presents reality from a different angle.
To investigate differences and relationships among different language editions, we develop mathematical and
statistical methods which treat the huge amount of information in Wikipedia,
excluding cultural preferences of the investigators.

Cultural bias or differences across Wikipedia
editions have been investigated in previous
research~\cite{ref3new,Callahan2011,Hecht2009,Hecht2010,Nemoto2011,Warnck-Wang2012,Massa2012}.
A special emphasis was devoted to persons
described in Wikipedia articles \cite{Callahan2011} and their ranking
\cite{Zhirov2010,Eom2013EPJB}.
Indeed, human knowledge, as well as Wikipedia
itself, was created by people who are the main actors of its development.
Thus it is rather natural to analyze a ranking of people
according to the Wikipedia hyper-link network of
citations between articles (see network data description below).
A cross-cultural study of biographical articles was presented
in~\cite{Aragon2012}, by building a network of interlinked
biographies. Another approach was proposed recently
in~\cite{Eom2013PLOS}: the difference in
importance of historical figures across Wikipedia language
editions is assessed on the basis of the global ranking of
Wikipedia articles about persons. This study, motivated by
the question ``Is an important person in a given culture also
important in other cultures?", showed that there
are strong entanglements and local biases of historical figures in Wikipedia.
Indeed, 
the results of the study show that
each Wikipedia edition favors persons belonging to the
same culture (language), but also that there are cross-Wikipedia top ranked persons,
who can be signs of entanglement between cultures.
These cross-language historical figures
can be used to
generate inter-culture networks demonstrating
interactions between cultures \cite{Eom2013PLOS}.
Such an approach provides us novel insights on cross-cultural differences
across Wikipedia editions. However, in~\cite{Eom2013PLOS} only 9 Wikipedia
editions, mainly languages spoken in European, have been considered.
Thus a broader set of language editions is needed to
offer a more complete view on a global scale.

We note that the analysis of persons' importance
via Wikipedia becomes more and more popular.
This is well visible from
the appearance of new recent studies for the English Wikipedia \cite{stonybrook}
and for multiple languages \cite{hidalgo}. The analysis of coverage of
researchers and academics
via Wikipedia is reported in \cite{tahaadded}.

Here we investigate interactions and skewness of cultures
with a broader perspective,
using global ranking of  articles about persons in 24
Wikipedia language editions.
According to Wikipedia~\cite{wikilist1}
these 24 languages cover 59 percent of world population.
Moreover, according to Wikipedia~\cite{wikilist2},
our selection of 24 language
editions covers the 68 percent of the total number of
30.9 millions of
Wikipedia articles in all 287 languages.
These 24 editions also cover languages which
played an important role in human history
including Western, Asian and Arabic cultures.

On the basis of this data set  we
analyze spatial, temporal, and gender skewness in Wikipedia by
analyzing birth place, birth date, and gender of the top ranked
historical figures in Wikipedia. We identified overall Western,
modern, and male skewness of important historical figures across
Wikipedia editions, a tendency towards local preference
(i.e. each
Wikipedia edition favors historical figures born in countries
speaking that edition's language), and the existence of global
historical figures who are highly ranked in most of Wikipedia
editions. We also constructed networks of cultures based on
cross-cultural historical figures to represent interactions between
cultures according to Wikipedia.

To obtain a unified ranking of historical figures
for all 24 Wikipedia editions,
we introduce an 
average ranking which gives us the top 100 persons of human history.
To assess the alignment of our ranking with previous work by historians,
we compare it with the Hart's list of the top 100 people who, according to him,
most influenced human history
\cite{hart}. We note that Hart ``ranked these 100 persons
in order of importance: that is, according to the total
amount of influence that each of them had on human history and on the
everyday lives of other human beings''.

\section*{Methods}
In this research, we consider each Wikipedia edition as a network
of articles. Each article corresponds to a node of the network and
hyperlinks between articles correspond to links of the network.
For a given network, we can define an adjacency matrix $A_{ij}$. If
there is a link (one or more) from node (article) $j$
to node (article)  $i$ then $A_{ij}=1$, otherwise, $A_{ij}=0$. The
out-degree $k_{out}(j)$ is the number of links from node $j$ to
other nodes and the in-degree $k_{in}(j)$ is the number of links
to node $j$ from other nodes. The links between articles are considered
only inside a given Wikipedia edition, there are no links
counted between editions. Thus each language edition is
analyzed independently from others by the Google matrix methods
described below. The transcriptions of names from English to the other 23
selected languages
are harvested from WikiData (http://dumps.wikimedia.org/wikidatawiki) and not directly
from the text of articles.

To rank the articles of a Wikipedia edition, we use two ranking algorithms based
on the articles network structure. Detailed descriptions
of these algorithms and their use for Wikipedia editions are given
in~\cite{Zhirov2010,2dmotor,Eom2013EPJB,frahm}. The methods used
here are described in~\cite{Eom2013PLOS}; we keep the same
notations.

\subsection*{Google matrix}
First we construct the matrix $S_{ij}$ of Markov transitions by
normalizing the sum of the elements in each column of $A$ to unity
($S_{ij}=A_{ij}/\sum_i A_{ij}$, $\sum_i S_{ij}=1$) and replacing
columns with zero elements by elements $1/N$ with $N$ being the
matrix size. Then the Google matrix is given by the relation
$G_{ij}=\alpha S_{ij}+(1-\alpha)/N$, where $\alpha$ is the damping
factor \cite{Meyer2006}. As in \cite{Eom2013PLOS} we use the conventional
value $\alpha=0.85$. It is known that the variation of $\alpha$
in a range $0.5 \leq \alpha < 0.95$ does not significantly
affect the probability distribution of ranks discussed below
(see e.g. \cite{Zhirov2010,Eom2013EPJB,Meyer2006}).

\subsection*{PageRank algorithm}
PageRank is a widely used algorithm to rank nodes in
a directed network. It was originally introduced for Google web
search engine to rank web pages of the World Wide Web based on the
idea of academic citations~\cite{Brin1998}. Currently PageRank is
used to rank nodes of network systems from scientific
papers~\cite{Chen2007} to social network services~\cite{Kwak2010},
world trade~\cite{wtrade} and biological
systems~\cite{Kandiah2013}. Here we briefly outline the iteration
method of PageRank computation. The PageRank vector $P(i,t)$ of a
node $i$ at iteration $t$ in a network with $N$ nodes is given by

\begin{equation}
P(i,t) = \sum_j G_{ij} P(j,t-1) = (1-\alpha)/N +
\alpha\sum_{j} A_{ij}P(j,t-1)/k_{out}(j)  \; . \label{eq2}
\end{equation}

 The stationary state $P(i)$ of $P(i,t)$ is the PageRank
of node $i$. More detailed information about the PageRank algorithm is
described in~\cite{Meyer2006}. Ordering all nodes by their
decreasing probability $P(i)$, we obtain the PageRank ranking
index $K(i)$. In qualitative terms, the PageRank probability of a node
is proportional to the number of incoming links
weighted according to their own probability. A random network surfer
spends on a given node a time given on average
by the PageRank probability.

\subsection*{CheiRank algorithm}
In a directed network, outgoing links can be as important as ingoing
links. In this sense, as a complementary to PageRank, the CheiRank
algorithm is defined and used in
~\cite{Chepelianskii2010,Zhirov2010,2dmotor}. The CheiRank vector
$P^*(i,t)$ of a node at iteration time $t$ is given by

\begin{equation}
P^*(i) = (1-\alpha)/N + \alpha\sum_{j} A_{ji}P^*(j)/k_{in}(j)
\label{eq3}
\end{equation}

Same as the case of PageRank, we consider the stationary state
$P^*(i)$ of $P^*(i,t)$ as the CheiRank probability of node $i$
with $\alpha=0.85$. High CheiRank nodes in the network have large
out-degree. Ordering all nodes by their decreasing probability
$P^*(i)$, we obtain the CheiRank ranking index $K^*(i)$.
The PageRank probability of an article is proportional to the number of incoming
links, while the CheiRank probability of an article is proportional to
the number of outgoing links. Thus a top PageRank article is important
since other articles refer to it, while a top CheiRank article
is highly connected because it refers to other articles.

\subsection*{2DRank algorithm}

PageRank and CheiRank algorithms focus only on in-degree and
out-degree of nodes, respectively. The 2DRank algorithm considers both types of
information simultaneously to rank nodes with a balanced point of
view in a directed network. Briefly speaking, nodes with both high
PageRank and CheiRank get high 2DRank ranking. Consider a
node $i$ which is $K_i$-th ranked by PageRank and ${K^*}_i$ ranked
by CheiRank. Then we can assign a secondary ranking
$K'_i=max\{K_i,{K^*}_i\}$ to the node. If $K'_i<K'_j$, then node
$j$ has lower 2DRank and vice versa. A detailed illustration and
description of this algorithm is given in \cite{Zhirov2010}.

We note that the studies reported in \cite{Eom2013PLOS}
show that the overlap between top  CheiRank persons of
different editions is rather small and due to that
the statistical accuracy of this data is not sufficient
for determining
interactions between different cultures for
the CheiRank list. Moreover,
CheiRank, based on outgoing links only, selects mainly
persons from such activity fields like
sports and arts
where the historical trace is not so important. Due to these
reasons we restrict our study to PageRank and 2DRank.
It can be also interesting to use other algorithms of ranking,
e.g. LeaderRank \cite{leaderrank},
but here we restrict ourselves to the methods
which we already tested, leaving investigation of other
raking methods for further studies.

\section*{Data preparation}
We consider 24 different language editions of Wikipedia:  English (EN), Dutch
(NL), German (DE), French (FR), Spanish (ES), Italian (IT),
Portuguese (PT), Greek (EL), Danish (DA), Swedish (SV), Polish
(PL), Hungarian (HU), Russian (RU), Hebrew (HE), Turkish (TR),
Arabic (AR), Persian (FA), Hindi (HI), Malaysian (MS), Thai (TH),
Vietnamese (VI), Chinese (ZH), Korean (KO), and Japanese (JA). The
Wikipedia data were collected in middle February 2013. The overview
summary of each Wikipedia is represented in Table ~\ref{table1}.

We understand that our selection of Wikipedia editions does not
represent a complete view of all the 287 languages of Wikipedia
editions. However, this selection covers most of the
largest language editions and allows us to perform
quantitative and statistical analysis of important historical
figures. Among the 20 largest editions
(counted by their size,
taken at the middle of 2014)
we have not considered the following
editions:  Waray-Waray, Cebuano, Ukrainian, Catalan,
Bokmal-Riksmal, and Finish.

First we ranked all the articles in a given Wikipedia edition by
PageRank and 2DRank algorithms, and selected biographical articles
about historical figures.
To identify biographical articles, we considered all articles belonging to
``Category:living people'',
or to ``Category:Deaths by year'' or ``Category:Birth by year''
or their subcategories in the English Wikipedia.
In this way, we obtained a list of about 1.1 million biographical articles.
We identified birth place,
birth date, and gender of each selected historical figure based on
DBpedia~\cite{DBpedia} or a manual inspection of
the corresponding Wikipedia biographical
article, when for the considered historical figure no DBpedia data were available.
We then started from the list of persons with their biographical article's title on the English Wikipedia, 
and found the corresponding titles in other language editions
using the inter-language links provided by WikiData.
Using the corresponding articles, identified by the inter-languages links
in different language editions,
we extracted the top 100 persons from the rankings of all Wikipedia articles of
each edition.
At the end, for each Wikipedia
edition and for each ranking algorithm, we have information about the top 100
historical figures with their corresponding name
in the English Wikipedia, their birth
place and date, and their gender. All 48 lists of the top 100
historical figures in PageRank and 2DRank
for the 24 Wikipedia editions and for the two ranking
algorithms are represented in~\cite{ourwikipage} and
Supporting Information (SI).
The original network data for each edition are available at
\cite{ourwikipage}.
The  automatic
extraction of persons from PageRank and 2DRank listings
of articles of each edition is performed by using
the above whole list of person names in all 24 editions.
This method implies a significantly higher recall
compared to the manual selection of persons from
the ranking list of articles for
each edition used in \cite{Eom2013PLOS}.

We attribute each of the 100 historical figures to a birth place at
the country level (actual country borders),
to a birth date in year, to a gender, and to a cultural group.
Historical figures are assigned to the countries currently
at the locations where they were born.
The cultural group of
historical figures is assigned by the most spoken language of
their birth place at the current country level. For example, if
someone was born in "Constantinople" in the ancient Roman era, since
the place is now Istanbul, Turkey, we assign her/his birth place
as "Turkey" and since Turkish is the most spoken language in
Turkey, we assign this person to the Turkish cultural group. If the
birth country does not belong to any of the 24 cultures (languages)
which we consider, we assign WR (world) as the culture of this
person. We would like to point out that although a culture can not
be defined only by language, we think that language is a
suitable first-approximation of culture. All lists of top 100
historical figures with their birth place, birth date, gender, and
cultural group for each Wikipedia edition and for each ranking
algorithm are represented in \cite{ourwikipage}. A part of this
information is also reported in SI.

To apply PageRank and 2DRank methods, we consider each edition as
the network of articles of the given edition connected by hyper-links among
the articles (see the details of ranking algorithms in Section Methods).
The full list of considered Wikipedia language editions is
given in Table~\ref{table1}. Table~\ref{table2} represents the top 10
historical figures by PageRank and 2DRank in the English Wikipedia.
Roughly speaking, top PageRank articles imply highly cited
articles in Wikipedia and top 2DRank articles imply
articles which are both  highly cited and highly citing
in Wikipedia. In total,
we identified 2400 top historical figures for each ranking
algorithm. However, since some historical figures such as {\it Jesus},
{\it Aristotle}, or {\it Napoleon}
appear in multiple Wikipedia editions,
we have 1045 unique top PageRank historical figures and
1616 unique top 2Drank historical figures.

We should note that the extraction of persons and their information from a
Wikipedia edition is
not an easy task even for the English edition, and much more complicated for
certain other language editions.
Therefore, the above automatic method
based on 1.1 million English names and their corresponding names
seems to us to be the most adequate approach.
Of course, it will miss people who do not have a biographical article
on the English Wikipedia.
Cross-checking investigation is done for Korean and Russian Wikipedia,
 which are native languages for two authors, by manually
selecting top 100 persons
from top lists of all articles ordered by PageRank and 2DRank in both
Wikipedia editions. We find
that our automatic search misses on average only 2 persons
from 100 top persons for these two editions
(the missed names are given in SI).
The errors appear due to transcription changes of
names or missing cases in our name-database
based on English Wikipedia.
For Western languages the number of errors is presumably
reduced since transcription remains close to English.
Based on the manual inspection for the Korean and the Russian Wikipedia, we expect that
the errors of our automatic recovery of the top people from the whole articles
ordered by PageRank and 2DRank are on a level of two percent.

We also note that our study is in compliance
with Wikipedia's Terms and Conditions.

\section*{Results}

Above we described the methods used for the extraction of the top 100
persons in the ranking list of each edition. Below we present
the obtained results describing the spatial, temporal and gender
distributions of top ranked historical figures. We also determine
the global and local persons and obtain the network of
cultures based on the ranking of persons from a given language by
other language editions of Wikipedia.

\subsection*{Spatial distribution}
The birth places of historical figures are attributed
to the country containing their geographical location of birth
according to the present geographical territories of
all world countries. The list of countries appeared
for the top 100 persons in all editions is given in Table~\ref{table3}.
We also attribute each country to one of the 24 languages
of the considered editions.
This attribution is done according to the language spoken by the largest
part of population in the given country.
Thus e.g. Belgium is attributed to Dutch (NL) since the majority
of the population speaks Dutch. If the main language of a country is
not among our 24 languages, then this country is attributed to
an additional section WR corresponding to the remaining world
(e.g. Ukraine, Norway are attributed to WR). If the birth place
of a person is not known, then it is also attributed to WR.
The choice of attribution of a person to a given country
in its current geographic territory, and as a result to a
certain language,
may have some fluctuations due to historical variations
of country borders (e.g. Immanuel Kant
was born in the current territory of Russia and hence is attributed
to Russian language). However, the number of such cases is small,
being on a level of 3.5 percent (see Section ``Network of cultures'' below).
We think that the way in which a link between person, language and country
is fixed by the birth place avoids much larger ambiguity
of attribution of a person according to the native language
which is not so easy to fix in an automatic manner.

\begin{figure*}[!ht]
\begin{center}
\includegraphics[width=0.7\columnwidth,angle=-90]{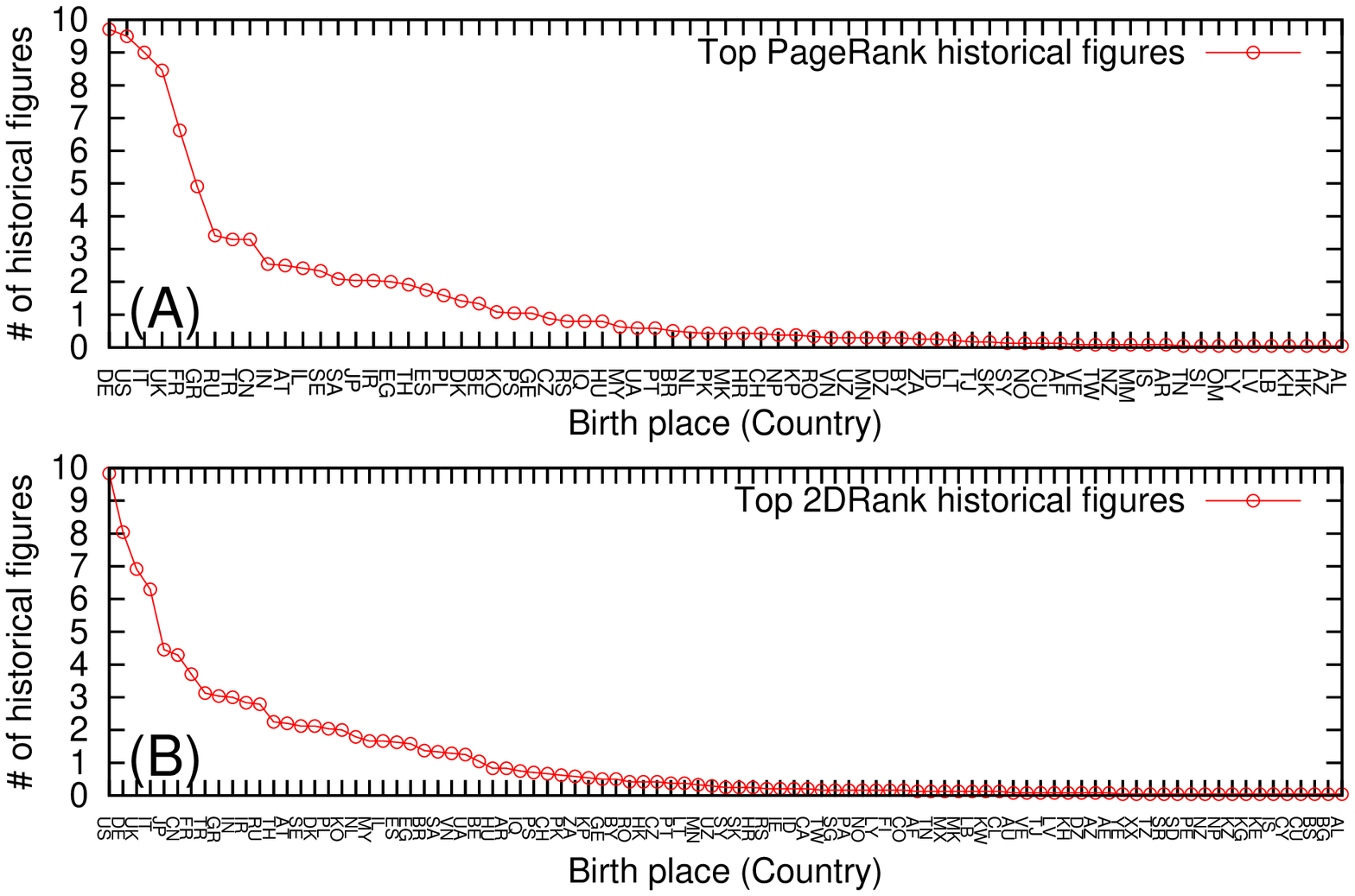}
\caption {\baselineskip 14pt Birth place distribution of top
historical figures averaged over 24 Wikipedia edition for (A)
PageRank historical figures (71 countries)
and (B) 2DRank historical figures (91 countries). Two
letter country codes are represented in Table~\ref{table3}.}
\label{fig1}\label{figure1}
\end{center}
\end{figure*}

The obtained spatial distribution of historical figures of
Wikipedia over countries is shown in Fig.~\ref{fig1}. This
averaged distribution gives the average number of top 100 persons
born in a specific country as birth place,
with averaging done over our 24 Wikipedia editions.
Thus an average over the 24 editions gives for
Germany (DE) approximately 9.7 persons in the top 100
of PageRank, being at the first position,
followed by USA with approximately 9.5 persons.
For 2DRank we have USA at the first position with an
average of 9.8 persons and Germany at the second
with an average of 8.0 persons.

Western (Europe and USA) skewed patterns are observed in
both top PageRank historical figures (Fig~\ref{fig1}. (A)) and top
2DRank historical figures (Fig.~\ref{fig1}. (B)). This Western
skewed pattern is remarkable since 11 Wikipedia editions of the 24
considered editions are not European language editions. Germany,
USA, Italy, UK and France are the top five birth places of top
PageRank historical figures among 71 countries. On the other
hand, USA, Germany, UK, Italy and Japan are top five birth
places of the top 2DRank historical figures among 91 countries.

\begin{figure*}[!ht]
\begin{center}
\includegraphics[width=0.92\columnwidth,angle=0]{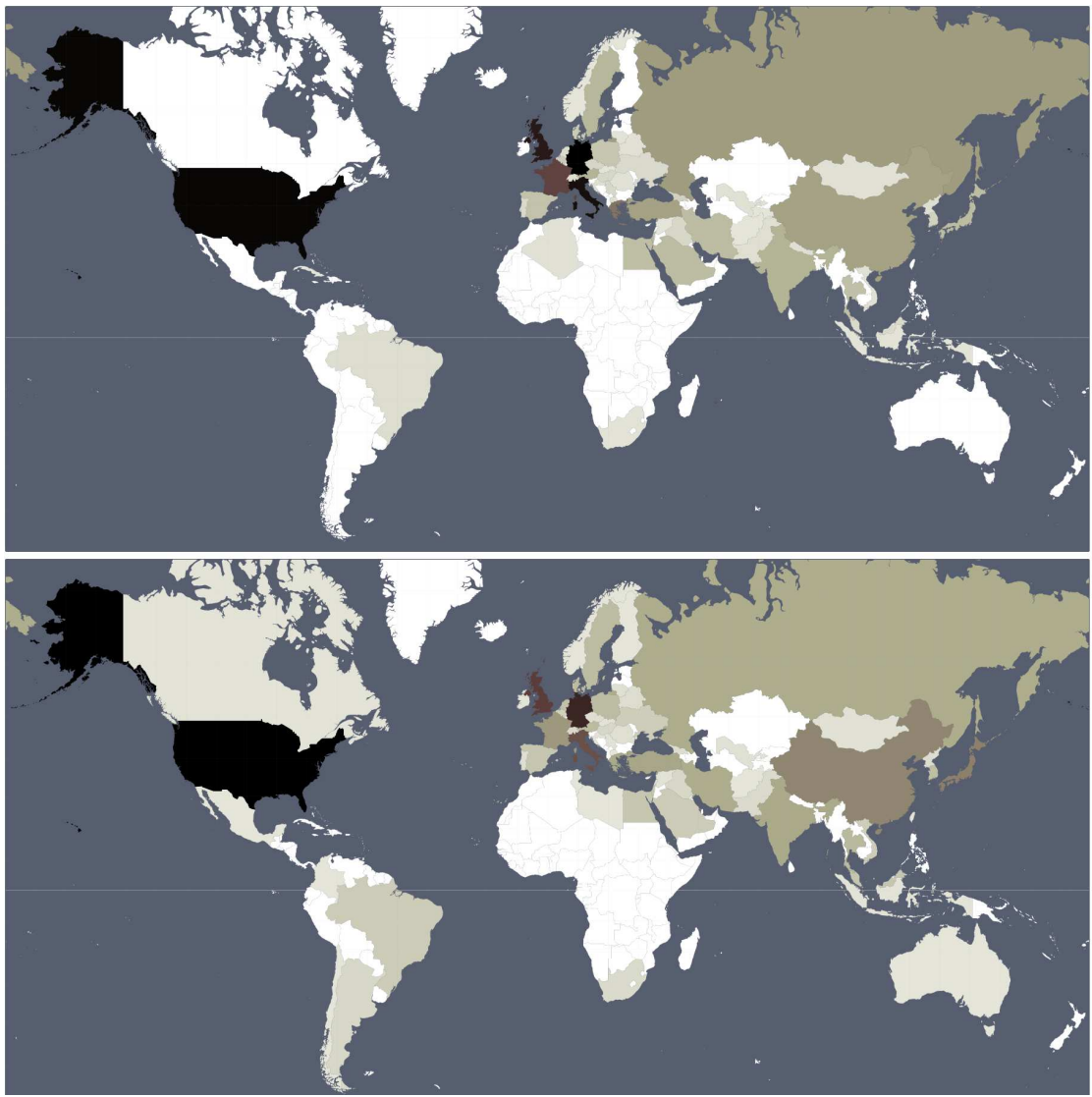}
\caption {\baselineskip 14pt
Sum of appearances of historical figures from a given country in the 24
lists of top 100 persons for PageRank (top panel) and 2DRank (bottom
panel). Color changes from zero (white) to maximum (black). Maximal
values are 233 appearances for Germany (top) and 236 for USA (bottom).
Values are proportional to the averages per country shown in
Fig.~\ref{fig1}.
}
\label{fig2}\label{figure2}
\end{center}
\end{figure*}

In Fig.~\ref{fig2} we show the world map of countries, where
color indicates the number of persons from a given country among
the $24 \times 100$ top persons for PageRank
and 2DRank. Additional figures showing these distributions
for different centuries are available at \cite{ourwikipage}.

\begin{figure*}[!ht]
\begin{center}
\includegraphics[width=0.92\columnwidth,angle=0]{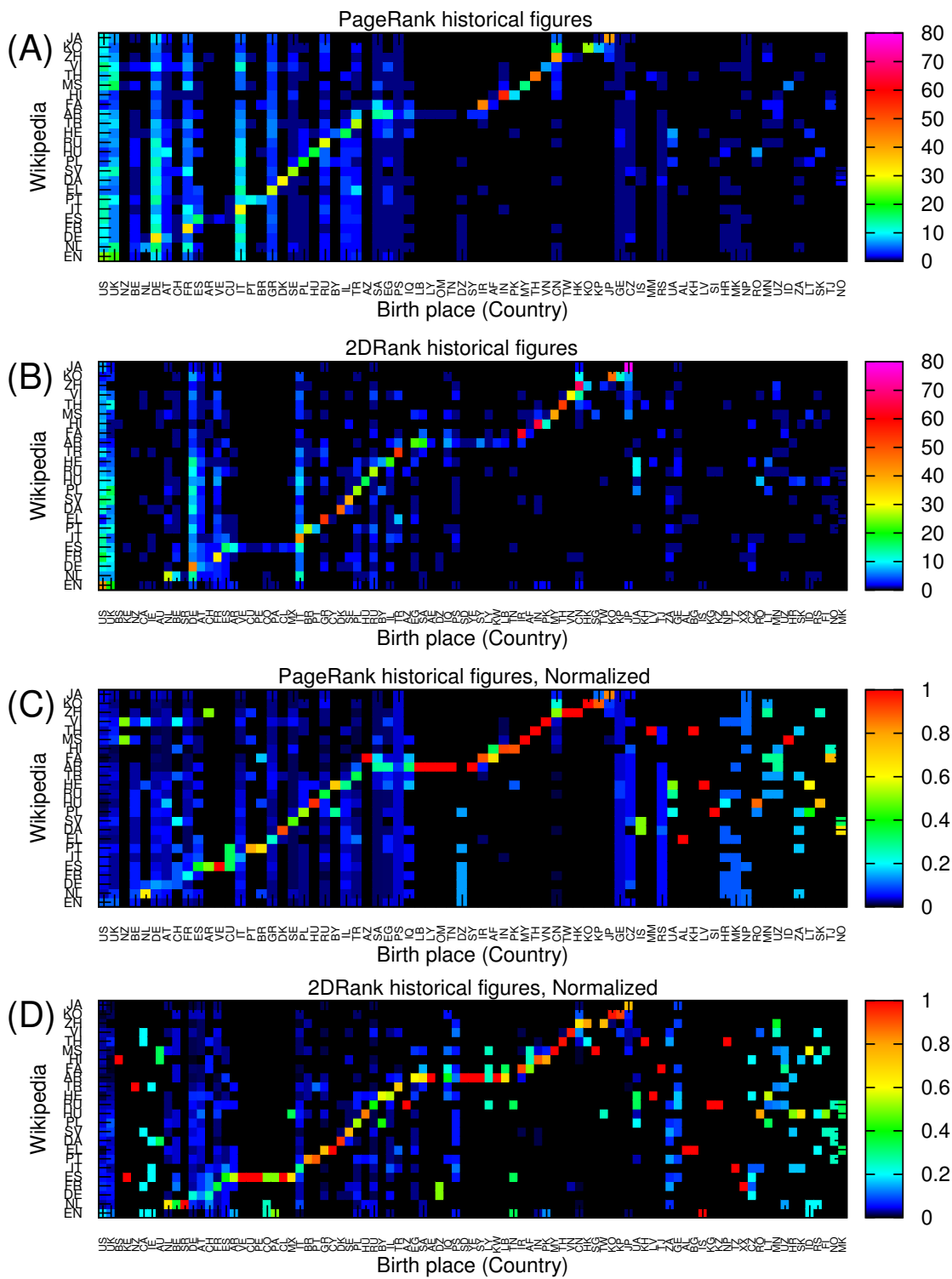}
\caption {\baselineskip 14pt Birth place distributions
over countries of top historical figures
from each Wikipedia edition; two letter country
codes are represented in Table~\ref{table3}. Panels:
(A) distributions of PageRank historical figures over 71 countries
for each Wikipedia edition; (B) distributions
of 2DRank historical figures
over 91 countries
for each Wikipedia edition; (C) column normalized birth
place distributions of PageRank historical figures
of panel (A); (D) column normalized birth place distributions
of 2DRank historical figures of panel (B).}
\label{fig3}\label{figure3}
\end{center}
\end{figure*}

We also observed local skewness in the spatial distribution of the top
historical figures for the PageRank (2DRank) ranking algorithm as
shown in Fig.~\ref{fig3}A (in Fig.~\ref{fig3}B). For example, 47
percent of the top PageRank historical figures in the English Wikipedia
were born in USA (25 percent) and UK (22 percent) and 56 percent of the
top historical figures in the Hindi Wikipedia were born in India.
A similar strong locality pattern of the top historical figures was
observed in our previous research~\cite{Eom2013PLOS}. However it
should be noted that in the previous study we considered the native
language of the top historical figure as a criterion of locality,
while in the current study we considered 'birth place' as criterion
of locality.

Regional skewness, the preferences of Wikipedia editions for
historical figures who were born in geographically or culturally
related countries, is also observed. For example, 18 (5) of the top
100 PageRank historical figures in the Korean (Japanese) Wikipedia
were born in China.
Also 9 of the top 100 PageRank historical figures in the Persian
Wikipedia were born in Saudi Arabia.
The distribution of top persons from each Wikipedia edition over
world countries is shown in Fig.~\ref{fig3}A and Fig.~\ref{fig3}B.
The countries on a horizontal axis are grouped
by clusters of corresponding language so that the links
inside a given culture (or language) become well visible.

To observe patterns in a better way
at low numbers of historical figures, we normalized each column of
Fig.~\ref{fig3}A and Fig.~\ref{fig3}B corresponding to a given country.
In this way we obtain a rescaled distribution with better visibility for
each birth country level as shown in Fig.~\ref{fig3}C and
Fig.~\ref{fig3}D, respectively.
We can observe a clear birth
pattern of top PageRank historical figures born in Lebanon, Libya,
Oman, and Tunisia in the case of the Arabic Wikipedia, and historical
figures born in N. Korea appearing not only in the Korean but
also in the Japanese Wikipedia.

In the case of the top 2DRank historical figures
shown in Fig.~\ref{fig3}B and Fig.~\ref{fig3}D, we observe overall
patterns of locality and regions
being similar to the case of
PageRank, but the locality
is stronger.

In short, we observed that most of the top historical figures in
Wikipedia were born in Western countries, but also that each edition
shows its own preference to the historical figures born in
countries which are closely related to the
corresponding language edition.

\subsection*{Temporal distribution}
\begin{figure*}[!ht]
\begin{center}
\includegraphics[width=0.9\columnwidth,angle=0]{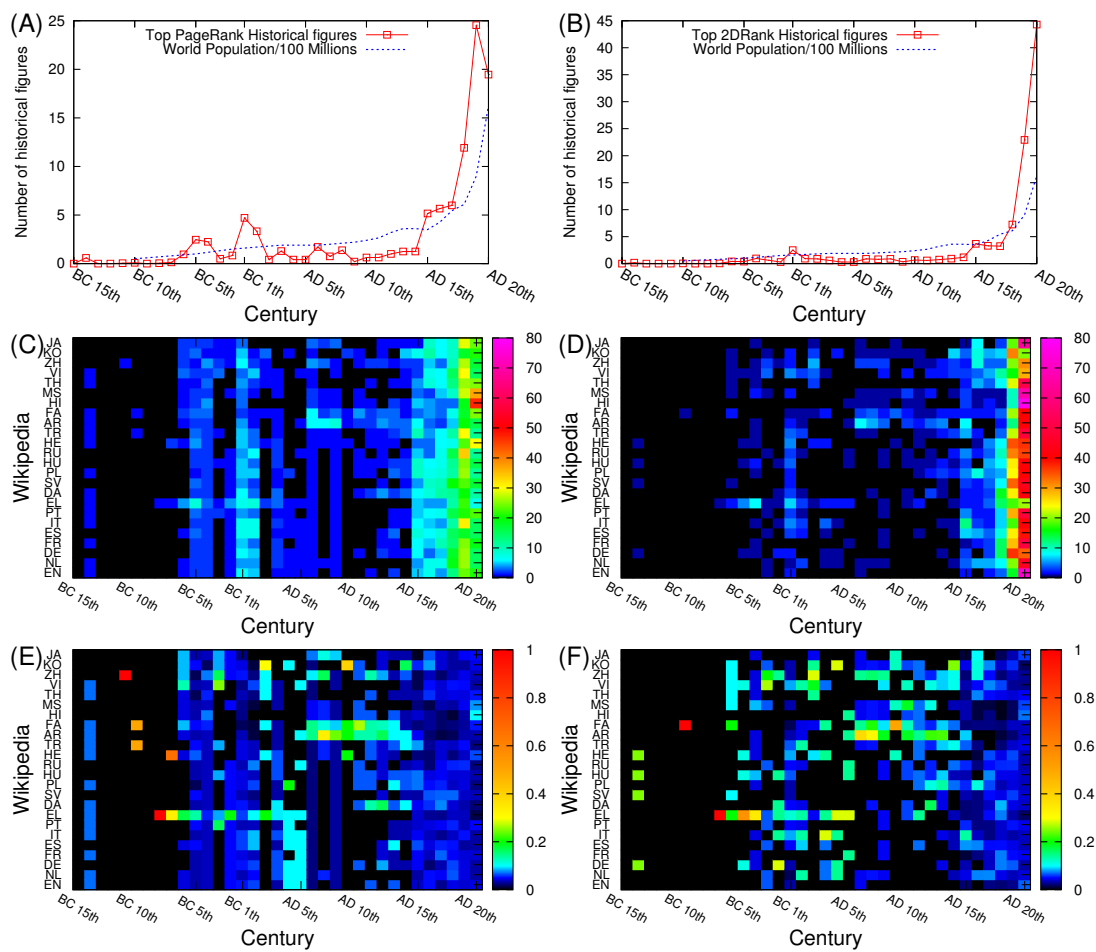}
\caption {\baselineskip 14pt Birth date distributions of top
historical figures. (A) Birth date distribution of PageRank
historical figures averaged over 24 Wikipedia editions (B) Birth
date distribution of 2DRank historical figures averaged over 24
Wikipedia editions (C) Birth date distributions of PageRank
historical figures for each Wikipedia edition. (D) Birth date
distributions of 2DRank historical figures for each Wikipedia
edition. (E) Column normalized birth date distributions of
PageRank historical figures for each Wikipedia edition. (F) Column
normalized birth date distributions of 2DRank historical figures
for each Wikipedia edition. } \label{fig4}\label{figure4}
\end{center}
\end{figure*}

The analysis of the temporal distribution of top historical figures
is done based on their birth dates. As shown in
Fig.~\ref{fig4}A for PageRank, most of historical figures were born
after the 17th century on average, which shows similar pattern with
world population growth \cite{populgrowth}.
However, there are some distinctive peaks
around BC 5th century and BC 1st century for the case of PageRank
because of Greek scholars ({\it Socrates, Plato,} and {\it Herodotus}), Roman
politicians ({\it Julius Caesar, Augustus}) and Christianity leaders
({\it Jesus, Paul the
Apostle,} and {\it Mary (mother of Jesus)}).
 We also
observe that the Arabic and the Persian Wikipedia have more historical figures
than Western language Wikipedia editions from AD 6th century to AD
12th century.
For the case of 2DRank
in Fig.~\ref{fig4}B,
there is only one small peak around BC 1C, which is also smaller than
the peak in the case of PageRank, and all the distribution
is dominated by a strong growth on the 20th
century.

The distributions of the top PageRank historical
figures over the 24 Wikipedia editions for each century
are shown in Fig.~\ref{fig4}C.
The same distribution, but normalized to unity over all editions
for each century, is shown in Fig.~\ref{fig4}E.
The Persian (FA) and the Arabic (AR) Wikipedia have
more historical figures than other language editions (in
particular European language editions) from the 6th to the 12th century
due to Islamic leaders and scholars. On the other hand, the Greek
Wikipedia has more historical figures in BC 5th century because
of Greek philosophers. Also most of western-southern European language
editions, including English, Dutch, German, French, Spanish,
Italian, Portuguese, and Greek, have more top historical figures
because they have {\it Augustine the Hippo} and
{\it Justinian I} in common.
Similar distributions obtained from 2DRank are shown in
 Fig.~\ref{fig4}D and Fig.~\ref{fig4}F respectively.

The data of Figs.~\ref{fig4}E,F clearly show
well pronounced patterns, corresponding to
strong interactions between cultures: from  BC 5th century
to  AD 15th century
for JA, KO, ZH, VI; from AD 6th century to
AD 12th century  for FA, AR;
and a common
 birth pattern in EN,EL,PT,IT,ES,DE,NL (Western European languages)
from  BC 5th century to  AD 6th century. In Fig.S1 in SI we
show distributions of historical figures over languages
according to their birth place. In this case the above patterns
become even more pronounced.

\begin{figure*}[!ht]
\begin{center}
\includegraphics[width=0.95\columnwidth,angle=0]{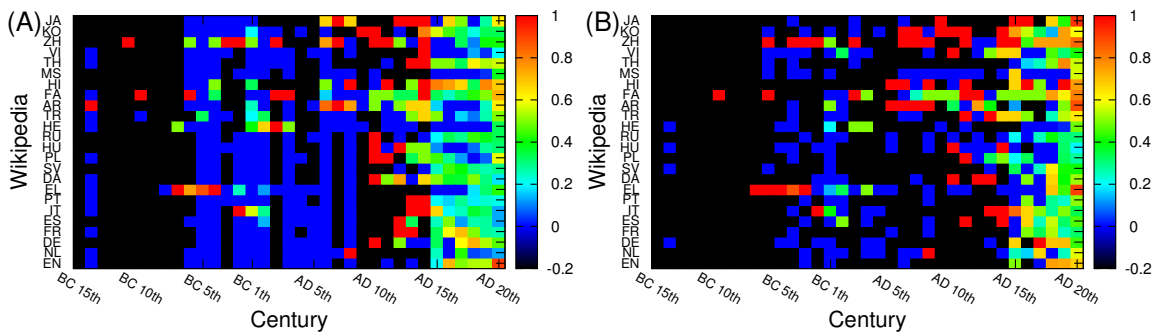}
\caption {\baselineskip 14pt The locality property of
cultures represented by the ratio $r_{L,C}=M_{L,C}/N_{L,C}$
for each edition $L$ and each century $C$. Here $M_{L,C}$ is the
number of historical figures born in countries attributed to
a given language edition $L$ at century $C$ and $N_{L,C}$ is the total
number of historical figures in a given edition at a given century,
regardless of language of their birth countries.
Black color (-0.2 in the color bars)
shows that there is no historical figure at all for a given edition
and century; blue (0 in the color bars) shows there there are
some historical figures but no same language historical figures.
Here (A) panel shows PageRank historical figures,
and  (B) panel shows 2DRank historical
figures.} \label{fig5}\label{figure5}
\end{center}
\end{figure*}

At a first glance from Figs.~\ref{fig4}E,F we observe
for persons born in AD 20th century
a significantly more homogeneous
distribution over cultures compared to early centuries.
However, as noted in \cite{Eom2013PLOS},
each Wikipedia edition favors
historical figures speaking the corresponding language. 
We investigate how this preference to
same-language historical figures changes in time. For this
analysis, we define two variables $M_{L,C}$ and $N_{L,C}$ for
a given language edition $L$ and a given century $C$.
Here $M_{L,C}$ is the number of historical figures born in all countries
being attributed to a given language $L$,
and $N_{L,C}$ is the total number of historical figures for a given century
$C$ and a given language edition $L$.
For example, among the 21 top PageRank
historical figures from the English Wikipedia, who were born in AD 20th
century, two historical figures (Pope John Paul II and Pope
Benedict XVI) were not born in English speaking countries.
Thus in this case $N_{EN,20}=21$ and $M_{EN,20}=19$. Fig~\ref{fig5}
represents the ratio $r_{L,C}=M_{L,C}/N_{L,C}$ for each edition and
each century. In ancient times (i.e. before AD 5th century),
most historical figures for each Wikipedia edition are not born in
the same language region except for the Greek, Italian, Hebrew, and
Chinese Wikipedia. However, after AD 5th century, the ratio of same
language historical figures is rising. Thus, in AD 20th century,
most Wikipedia editions have significant numbers of historical
figures born in countries speaking the corresponding language.
For PageRank persons and AD 20th century,
we find that the English edition has the largest fraction of its own
language, followed
by Arabic and Persian editions while other editions have significantly
large connections with other cultures.
For the English edition this is related to a significant
number of USA presidents appearing in the top 100 list
(see \cite{Zhirov2010,Eom2013EPJB}).
For 2DRank persons the largest fractions
were found for Greek, Arabic, Chinese and Japanese cultures.
These data show that even in age of globalization there is a significant
dominance of local historical figures for certain cultures.

\subsection*{Gender distribution}
\begin{figure*}[!ht]
\begin{center}
\includegraphics[width=0.65\columnwidth,angle=-90]{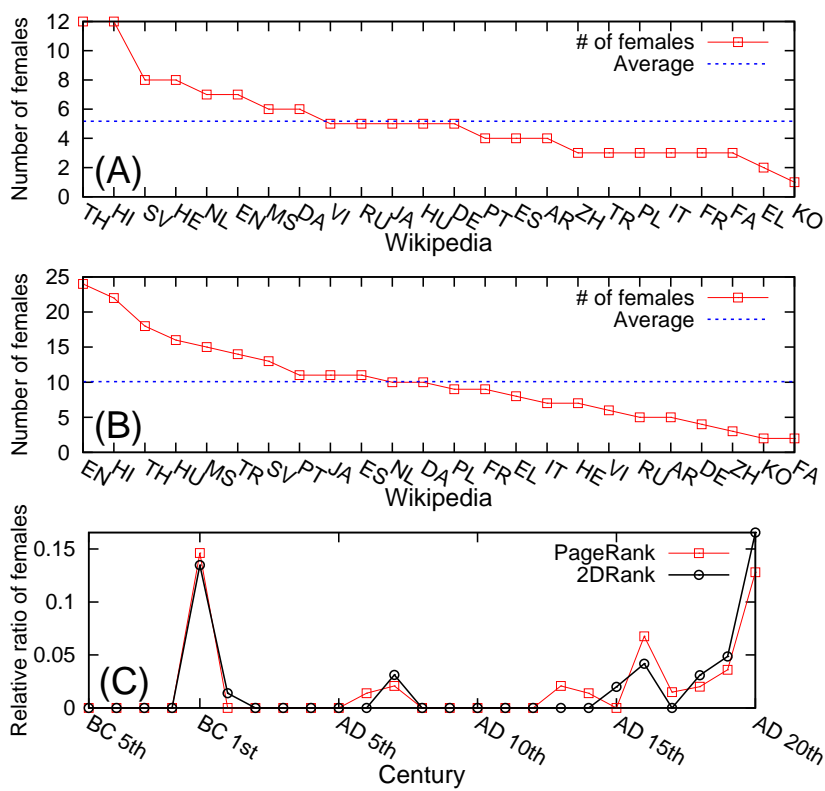}
\caption {\baselineskip 14pt Number of females of top historical
figures from each Wikipedia edition (A) Top PageRank historical
figures (B) Top 2DRank historical figures. (C) The average female
ratio of historical figures in given centuries across 24 Wikipedia
editions. } \label{fig6}\label{figure6}
\end{center}
\end{figure*}

From the gender distributions of historical figures, we observe
a strong male-skewed pattern across many Wikipedia editions regardless of the
ranking algorithm.
On average, $5.2 (10.1)$ female historical figures are observed among the 100 top
PageRank (2DRank) persons for each Wikipedia edition.
Fig.~\ref{fig6} shows the number of top female historical figures
for each Wikipedia edition. Thai, Hindi, Swedish, and Hebrew have
more female historical figures than the average over our 24 editions
in the case of PageRank
On the other hand, the Greek and the Korean versions have
a lower number of females than the average. In the case of 2DRank,
English, Hindi, Thai, and Hungarian Wikipedia have more females
than the average while German, Chinese, Korean, and Persian
Wikipedia have less females than the average.
In short, the top historical figures in Wikipedia are quite male-skewed. This is
not surprising since females had little chance to be
historical figures for most of human history. We compare the
gender skewness 
to other cases such as the number of
female editors in Wikipedia (9 percent) in 2011~\cite{Wikipedians}
and the share of women in parliaments, which was 18.7 percent in
2012 by UN Statistics and indicators on women and
men~\cite{UN_Women2012}, the male skewness
for the PageRank list is stronger in the
contents of Wikipedia \cite{Lam2011}.
 However, the ratio of females among the top
historical figures is growing by time as shown in Fig.~\ref{fig6}
C. It is notable that the peak in Fig.~\ref{fig6}C at BC 1st is
due to "Mary (mother of Jesus)". In the 20th century
2DRank gives a larger percentage of women
compared to PageRank. This is due to the fact that
2DRank has a larger fraction of singers and artists
comparing to PageRank (see \cite{Zhirov2010,Eom2013EPJB})
and that the fraction of women in these
fields of activity is larger.

\subsection*{Global historical figures}
Above we analyzed  how top historical figures in Wikipedia are
distributed in terms of space, time, and gender. Now we identify how
these top historical figures are distributed in each Wikipedia
edition and which are global historical figures. According to previous
research~\cite{Eom2013PLOS},
there are some global historical figures who are recognized as important
historical figures across Wikipedia editions.
We identify global historical figures based on the ranking score for a
given person determined by her number of appearances
and ranking index over our 24 Wikipedia editions.

Following ~\cite{Eom2013PLOS}, the ranking score $\Theta_{P,A}$
of a historical figure $P$ is given by

\begin{equation}
\Theta_{P,A} = \sum_{E} (101-R_{P,E,A}) \label{eq1}
\end{equation}

Here $R_{P,E,A}$ is the ranking of a historical
figure $P$ in Wikipedia edition $E$ by ranking algorithm $A$.
According to this
definition, a historical figure who appears more often in the
lists of top historical figures for the given 24 Wikipedia editions or
has higher ranking in the lists gets a higher ranking score.
Table~\ref{table4} represents the top 10 global historical
figures for PageRank and 2DRank. {\it Carl Linnaeus}
is the 1st global
historical figure by PageRank followed by {\it Jesus, Aristotle}.
{\it Adolf Hitler} is the 1st global
historical figure by 2DRank followed by {\it Michael Jackson, Madonna (entertainer)}.
On the other hand, the lists of the top 10 local historical figures ordered by
our ranking score for each language are represented in
SI Tables S1-S25 and \cite{ourwikipage}.

The reason for a somewhat unexpected PageRank leader  {\it Carl Linnaeus}
is related to the fact that he laid the foundations for the modern biological naming scheme
so that plenty of articles about animals, insects and plants
point to the Wikipedia
article about him, which strongly increases the PageRank probability.
This happens for all 24 languages where {\it Carl Linnaeus} always
appears on high positions since articles about animals and plants are an
important fraction of Wikipedia.
Even if in a given language
the top persons are often politicians (e.g. {\it Napoleon, Barak Obama} at $K=1,2$ in EN),
these politicians have mainly local importance and
are not highly ranked in other languages
(e.g. in ZH {\it Carl Linnaeus} is at $K=1$,
{\it Napoleon} at $K=3$ and {\it Barak Obama} is at $K=24$).
As a result when the global contribution is counted over all $24$
languages {\it Carl Linnaeus} appears on the top PageRank position.

\begin{figure*}[!ht]
\begin{center}
\includegraphics[width=0.9\columnwidth,angle=0]{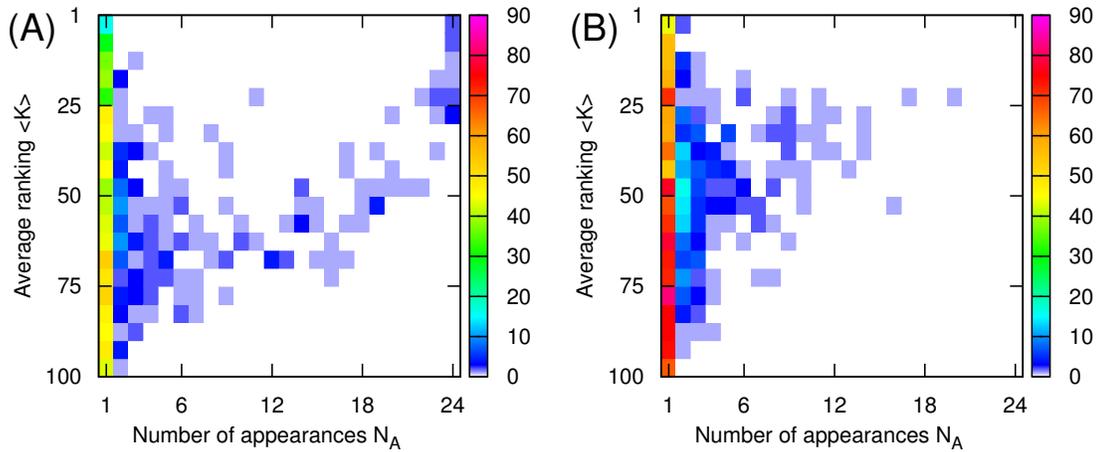}
\caption {\baselineskip 14pt
The distribution of 1045 top PageRank persons (A)
and 1616 top 2DRank persons (B) as a function of number
of appearances $N_A$ of a given person and
the rank $\langle K \rangle$ of this
person averaged over Wikipedia editions where this person
appeared.}
\label{fig7}\label{figure7}
\end{center}
\end{figure*}

Our analysis suggests that there might be three groups of historical
figures. Fig.~\ref{fig7} shows
these three groups of top PageRank historical figures in Wikipedia: (i)
global historical figures
who appear in most of Wikipedia editions ($N_A \geq 18$) and are highly
ranked ($\langle K \rangle  \leq 50$) for each Wikipedia such as
Carl Linnaeus, Plato, Jesus, and Napoleon (Right-Top of the Fig. 7A); (ii)
local-highly ranked
historical figures who appear in a few Wikipedia editions ($N_A<18$)  but
are highly ranked ( $\langle K \rangle  \leq 50$ ) in the
Wikipedia editions in which they appear,
such as Tycho Brahe, Sejong the Great, and Sun Yat-sen
(Left-Top of the
Fig. 7A); (iii) locally-low ranked historical figures who appear in a few
Wikipedia editions ($N_A<18$) and
who are not highly ranked ($\langle K \rangle > 50$). Here $N_A$ is the
number of appearances in different Wikipedia editions for a given person
and $\langle K \rangle$
is the average ranking of the given persons across Wikipedia
editions for each ranking algorithm.
In the case of 2DRank historical figures, due to the absence of global
historical figures, most of them belong to two types of local historical
figures (i.e. local-highly
ranked or local-lowly ranked).


Following ranking of persons via $\Theta_{P,A}$
we determine also the top global female historical figures,
presented in Table~\ref{table5} for PageRank and 2DRank
persons. The full lists of global female figures
are available at \cite{ourwikipage} (63 and 165 names
for PageRank and 2DRank).

The comparison of our 100 global historical figures
with the top 100 from Hart's list ~\cite{hart} gives an overlap
of 43 persons for PageRank and 26 persons for 2DRank.
We note that for the top 100 from the English Wikipedia we obtain
a lower overlap of 37 (PageRank) and 4 (2DRank) persons.
Among all editions the highest overlaps with the Hart list
are 42 (VI), 37 (EN,ES,PT,TR) and 33 (IT), 32 (DE), 31 (FR) for PageRank;
while for 2DRank we find 18 (EL) and 17 (VI).
We give the overlap numbers for all editions at \cite{ourwikipage}.
This shows that the consideration of 24 editions
provides us the global list of the top 100 persons
with a more balanced selection of top historical figures.
Our overlap of the top 100 global historical figures
by PageRank with the top 100 people
from Pantheon MIT ranking list \cite{hidalgo}
is  44 percent, while the overlap of this Pantheon list with Hart's list is
43 percent.
We note that the Pantheon method is significantly based on  a number of
page views while our approach is based on the network structure
of the whole Wikipedia network.
The top 100 persons from \cite{stonybrook}
are not publicly available  but nevertheless
we present the overlaps between the top 100 persons from the
lists of Hart, Pantheon, Stony-Brook and our global
PageRank and 2DRank lists in SI Figs.S2,S3 (we received the Stony-Brook list
as a private message from the authors of \cite{stonybrook}).
We have an average overlap between the 4 methods on a level of 40 percent
(2DRank is on average lower by a few percent),
we find a larger overlap between our PageRank list and the Stony-Brook
list since the Stony-Brook method, applied only for the
English Wikipedia, is
significantly based on PageRank.

We also compared the distributions of our global top 100 persons of PageRank and
2DRank with the distribution of Hart's top 100 over centuries
and over 24 languages
with the additional WR category (see Fig.S4 in SI).
We find that these 3 distributions
have very similar shapes. Thus the largest number
of persons appears in centuries
AD 18th, 19th, 20th for the 3 distributions. Among languages, the main peaks
for the 3 distributions appear for EN, DE, IT, EL, AR, ZH.
The deviations from Hart's
distribution are larger  for the 2DRank list. Thus the comparison of
distributions over centuries and languages shows that
the PageRank list has not only
a strong overlap with the Hart list in the number of persons
but that  they also have very similar
statistical distributions of the top 100 persons
over centuries and languages.

The overlap of the top 100 global persons
found here with the previous study~\cite{Eom2013PLOS}
gives 54 and 47 percent for PageRank and 2DRank lists, respectively.
However, we note that the global list in \cite{Eom2013PLOS}
was obtained from the top 30 persons in each edition
while here we use the top 100 persons.

It is interesting to note that for the top 100 PageRank universities
from the English Wikipedia edition
the overlap
with Shanghai top 100 list of universities is
on a even higher level of 75 percent~\cite{Zhirov2010}.

Finally, we note that the ranking of historical figures
using the whole PageRank (or 2DRank) list
of all Wikipedia articles of a given edition
provides a more stable approach compared to the network
of biographical articles used in~\cite{Aragon2012}.
Indeed, the number of nodes and links in such a biographical network
is significantly smaller compared to the whole
network of Wikipedia articles and thus the fluctuations become rather large.
For example, from the biographical network
of the Russian edition one finds as the top person {\it Napoleon III}
(and even not {\it Napoleon I})~\cite{Aragon2012},
who has a rather low importance for
Russia. In contrast to that the present study gives us the top PageRank
historical figure of the Russian edition to be
{\it Peter the Great}, that has much more historical grounds.
In a similar way for FR the results of \cite{Aragon2012}
give at the first position
{\it Adolf Hitler}, that is rather strange for the French
culture, while we find a natural result {\it Napoleon}.

\subsection*{Network of cultures}

We consider the selected top persons from each
Wikipedia edition as important historical figures recognized by
people who speak the language of that
Wikipedia edition. Therefore, if a top person from
a language edition $A$ appears in another edition
$B$, then we can consider this as a 'cultural' influence from culture $A$
to $B$. Here we consider each language as a proxy for a cultural
group and assign each historical figure to one of these cultural
groups based on the most spoken language of her/his birth place at the
country level. For example, {\it Adolf Hitler} was born in modern
Austria and since German language is the most spoken language in
Austria, he is considered as a German historical figure in our
analysis. This method may lead to some
misguiding results due to discrepancy
between territories of country and cultures,
e.g. {\it Jesus}  was
born in the modern State of Palestine (Bethlehem), which is an Arabic speaking
country. Thus {\it Jesus} is from the Arabic culture in our analysis
while usually one would say that he belongs to the Hebrew culture.
Other similar examples we find are:
{\it Charlemagne}  (Belgium - Dutch),
{\it Immanuel Kant} (Russia - Russian, while usually he is attributed to DE),
{\it Moses} (Egypt - Arabic),
{\it Catherine the Great} (Poland - Polish, while usually she would be attributed to DE or RU).

In total there are such 36 cases from the global PageRank list of
1045 names (these 36 names are given in SI).
However, in our knowledge, the birth place
is the best way to assign a given historical figure to
a certain cultural background computationally and
systematically and with the data we have available.
In total we have only about 3.4 percent
of cases which can be discussed and where a native speaking language
can be a better indicator of belonging to a given culture.
For the global 2DRank list of 1616 names we identified
53 similar cases where an attribution to a culture
via a native language
or a birth place could be discussed (about 3.3 percent).
These 53 names are given in SI.
About half of such cases are linked with birth places in
ancient Russian Empire where people
from Belarus, Litvania and Ukraine moved to RU, IL, PL, WR.
However, the percentage of such cases is
small and the corresponding errors also remain small.

Based on the above assumption and following the approach
developed in \cite{Eom2013PLOS}, we construct two weighted
networks of cultures (or language groups) based on the top PageRank
historical figures and top 2DRank historical figures respectively.
Each culture (i.e. language) is represented
as a node of the network, and the weight
of a directed link from culture $A$ to culture $B$ is given by the number of
historical figures belonging to culture $B$ (e.g. French)
appearing in the list of top 100 historical figures for a given
culture $A$ (e.g. English).
The persons in a given edition, belonging
to the language of the edition, are not taken into account
since they do not create links between cultures.
In Table~\ref{table6} we give the number of such persons
for each language. This table also gives the number of persons
of a given language among the top 100 persons of
the global PageRank and 2DRank listings.

For example, there are 5 French historical figures among the top 100 PageRank
historical figures of the English Wikipedia,
so we can assign weight 5 to the link from English to
French. Fig.~\ref{fig8}A and Fig.~\ref{fig8}B represent the
constructed networks of cultures defined by appearances of the top
PageRank historical figures and top 2DRank historical figures,
respectively.
In total we have two networks with 25 nodes which include our 24 editions
and an additional node WR for all the other world cultures.

The Google matrix $G_{ij}$
for each network is constructed following the standard rules
described in \cite{Eom2013PLOS} and in the Methods Section.
In a standard way we determine the PageRank
index $K$ and the CheiRank index $K^*$
that order all cultures according to decreasing
PageRank and CheiRank probabilities (see Methods and Fig.S5. in SI).
The structure of matrix elements $G_{KK'}$ is
shown in Fig.~\ref{fig9}.

To identify which cultures (or language groups) are more
influential than others, we calculated PageRank and CheiRank of
the constructed networks of cultures by considering link weights.
Briefly speaking, a culture has high PageRank (CheiRank) if
it has many ingoing (outgoing) links
from (to) other cultures (see Methods). The distribution of cultures
on a PageRank-CheiRank plane is shown in Fig.~\ref{fig10}.
In both cases of PageRank and
2DRank historical figures, historical figures of English culture
(i.e. born in English language spoken countries) are the most
influential (highest PageRank) and German culture is the second one
(Fig.~\ref{fig10}A,B).
Here we consider the historical figures for the whole
range of centuries.
Fig.~\ref{fig10} represents the detailed features of how each culture
is located on the plane of PageRank ranking $K$ and CheiRank
ranking $K*$ based on the top PageRank historical
figures (Fig.~\ref{fig10}A) and top 2DRank historical figures
(Fig.~\ref{fig10}B). Here $K$ indicates the ranking of a given
culture ordered by how many of its own top historical figures appear
in other Wikipedia editions, and $K^*$ indicates the ranking of a given
culture according to how many of the top historical figures in the
considered culture are from other cultures. As described above,
English is on ($K=1$, $K^*=19$) and German is on ($K=2$, $K^*=21$)
in the case of PageRank historical figures (Fig.~\ref{fig10}A). In
the case of 2DRank historical figures, English is on ($K=1$,
$K^*=14$) and German is on ($K=2$, $K^*=9$).

It is important to note that there is a significant difference
compared to the previous study \cite{Eom2013PLOS}:
there, only 9 editions had been considered and the top
positions were attributed to the world node WR
which captured a significant fraction of the top persons.
This indicated that 9 editions are not sufficient to
cover the whole world. Now for 24 editions
we see that the importance of the world node WR
is much lower (it moves from $K=1$ for
9 editions \cite{Eom2013PLOS} to
$K=4$ and $3$ in Fig.~\ref{fig10}A and Fig.~\ref{fig10}B).
Thus our 24 editions cover the majority the world.
Still it would be desirable to add a few additional
editions (e.g. Ukraine, Baltic Republics, Serbia etc.)
to fill certain gaps.

It is interesting to note that the ranking plane
of cultures $(K,K^*)$ changes significantly in time.
Indeed, if we take into account only persons born
before the 19th century then the
ranking is modified with
EN going to 4th (Fig.~\ref{fig10}C for PageRank figures)
and 6th position (Fig.~\ref{fig10}C for 2DRank figures)
while the top positions are taken by IT, DE, FR and
DE, IT, AR, respectively.

At the same time,
we may also argue that for cultures it is important
not only to be cited but also to be communicative
with other cultures.
To characterize communicative properties of nodes on the network
of cultures shown in Fig.~\ref{fig8}
we use again the concepts of PageRank, CheiRank and 2DRank
for these networks as described in Methods and \cite{Eom2013PLOS}.
Thus, for the network of cultures of Fig.~\ref{fig8},
the 2DRank index
of cultures highlights their influence in a more
balanced way taking into account their importance
(incoming links) and communicative (outgoing links)
properties in a balanced manner.

Thus we find for all centuries
at the top positions
Greek, Turkish and Arabic (for PageRank persons)
and French, Russian and Arabic (for 2DRank persons).
For historical figures before the 19th century,
we find respectively
Arabic, Turkish and Greek (for PageRank)
and Arabic, Greek and Hebrew (for 2DRank).
The high position of Turkish is due to its
close links both with Greek culture in ancient times
and with Arabic culture in more recent times.
We see also that with time the positions
of Greek in 2DRank improves due to a global
improved ranking of Western cultures closely
connected with Greece.

\section*{Discussion}
By investigating birth place, birth date, and gender of important
historical figures determined by the network structure of Wikipedia,
we identified spatial, temporal, and gender skewness in Wikipedia.
Our analysis shows that the most important historical figures across
Wikipedia language editions were born in Western countries
after the 17th century, and are male. Also, each Wikipedia
edition highlights local figures so that most of its own historical
figures are born in the countries which use the language
of the edition. The emergence of such pronounced accent
to local figures seems to be natural since there are more links and
interactions within one culture.
This is also visible
from to the fact that in many editions the main country for the given
language is at the first PageRank position among all
articles (e.g. Russia in RU edition) \cite{Eom2013PLOS}.
Despite such a locality feature,
there are also global historical figures who appear in
most of the considered Wikipedia editions with very high rankings. Based
on the cross-cultural historical figures, who appear in multiple
editions, we can construct a network of cultures which describes
interactions and entanglement between cultures.

It is very difficult to
describe history in an objective way
and due to that it was argued that history is
"an unending dialogue between the past and
present"~\cite{Carr1961}. In a similar way we can say that
history is an unending
dialogue between different cultural groups.
We use a computational and data mining approach, based on rank vectors
of the Google matrix of Wikipedia, to perform a statistical analysis of
interactions and entanglement of cultures. We find that this approach
can be used for selecting the most influential historical figures
through an analysis of collectively generated links between articles
on Wikipedia. Our results are coherent with studies conducted by
historians \cite{hart}, with an overlap of 43\% of important historical
figures. Thus, such a mathematical analysis of local and global
historical figures can be a useful step towards the understanding of
local and global history and interactions of world cultures. Our
approach has some limitations, mainly caused by the data source and by
the difficulty of defining culture boundaries across centuries.
The ongoing improvement of structured content in Wikipedia through the
WikiData project, eventually in conjunction with additional manual
annotation, should allow to deal with these limitations. Furthermore, it
would be useful to perform comparisons with other approaches to measure
the interactions of cultures, such as the analysis of language crossings
of multilingual users \cite{hale}.

Influence of  digital media on information dissemination
and social collective opinions among the public is growing fast.
 Our research across
Wikipedia language editions suggests
a rigorous mathematical way, based on Markov chains and Google matrix,
for the identification of important historical figures
and for the analysis of interactions of cultures
at different historical periods and in different
world regions. We think that  a further extension of this approach
to a larger number of Wikipedia editions will provide
a more detailed and balanced analysis of interactions of
world cultures.


\section*{Acknowledgments}

This research is supported in part by the EC FET Open project
``New tools and algorithms for directed network analysis''
(NADINE $No$ 288956).

\section*{Supporting Information}

\begin{figure*}[!ht]
\begin{center}
\includegraphics[width=0.95\columnwidth,angle=0]{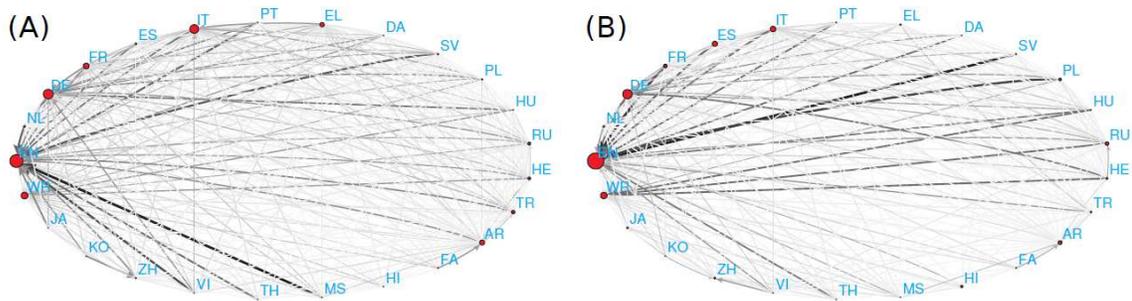}
\caption {\baselineskip 14pt Network of cultures obtained from 24
Wikipedia languages and the remaining world (WR) consider (A) top
PageRank historical figures and (B) 2DRank historical figures. The
link width and darkness are proportional to a number of foreign
historical figures quoted in top 100 of a given culture, the link
direction goes from a given culture to cultures of quoted foreign
historical figures, links inside cultures are not considered.
The size of nodes is proportional to their PageRank.}
\label{fig8}\label{figure8}
\end{center}
\end{figure*}

\begin{figure*}[!ht]
\begin{center}
\includegraphics[width=0.95\columnwidth,angle=0]{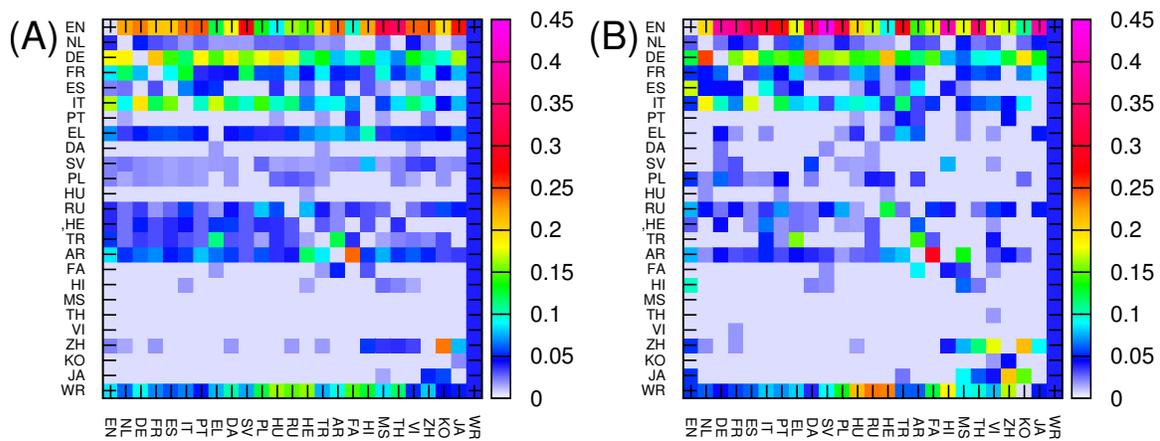}
\caption {\baselineskip 14pt Google matrix of network of cultures
shown in Fig~\ref{fig8} respectively. The matrix elements $G_{ij}$
are shown by color with damping factor $\alpha=0.85$.}
\label{fig9}\label{figure9}
\end{center}
\end{figure*}

\begin{figure*}[!ht]
\begin{center}
\includegraphics[width=0.7\columnwidth,angle=-90]{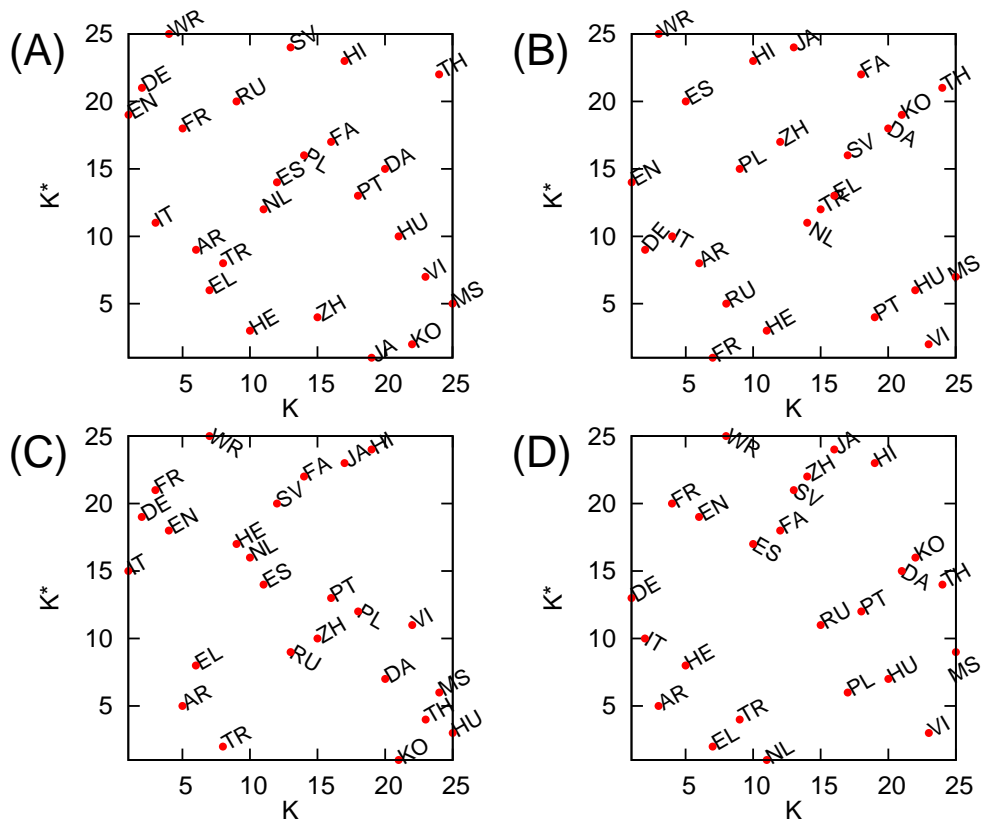}
\caption {\baselineskip 14pt PageRank ranking versus CheiRank
ranking plane of cultures with corresponding indexes $K$ and $K^*$
obtained from the network of cultures based on (A) all PageRank
historical figures, (B) all 2DRank historical figures, (C)
PageRank historical figure born before AD 19th century, and (D)
2DRank historical figure born before AD 19th century,
respectively.} \label{fig10}\label{figure10}
\end{center}
\end{figure*}

Supporting Information file
presents Figure S1 with PageRank and CheiRank probabilities
for networks of cultures of Fig.~\ref{fig8},
lists of top 100 global PageRank and 2DRank names;
Tables S1-S25 of top 10 names of given language
and remained world from the global PageRank and 2DRank
ranking lists of persons ordered by the score $\Theta_{P,A}$
of Eq.(\ref{eq1}). Additional lists of all 100 ranked names
for all 24 Wikipedia editions and corresponding network link data
for each edition are given at \cite{ourwikipage}.

\newpage$\phantom{.}$

\newpage$\phantom{.}$

\begin{table*}[!ht]
\caption{Wikipedia hyperlink networks from the 24 considered language editions.
Here $N_a$ is the number of articles. Wikipedia data were collected in middle February 2013. }
\begin{center}
\resizebox{12cm}{!}{
\begin{tabular}{|c|c|c|c|c|c|}
  \hline
  Edition & Language & $N_{a}$ & Edition & Language & $N_{a}$\\
  \hline
EN & English    &4212493 & RU & Russian    &966284 \\
NL & Dutch      &1144615 & HE & Hebrew     &144959 \\
DE & German     &1532978 & TR & Turkish    &206311 \\
FR & French     &1352825 & AR & Arabic     &203328 \\
ES & Spanish    &974025  & FA & Persian    &295696 \\
IT & Italian    &1017953 & HI & Hindi      &96869  \\
PT & Portuguese &758227  & MS & Malaysian  &180886 \\
EL & Greek      &82563   & TH & Thai       &78953  \\
DA & Danish     &175228  & VI & Vietnamese &594089 \\
SV & Swedish    &780872  & ZH & Chinese    &663485 \\
PL & Polish     &949153  & KO & Korean     &231959 \\
HU & Hungarian  &235212  & JA & Japanese   &852087 \\
  \hline
\end{tabular}}
\end{center}
\label{table1}
\end{table*}

\newpage$\phantom{.}$
\begin{table*}[!ht]
\caption{List of top persons by PageRank and 2DRank for
the English Wikipedia. All names are represented by article
titles in the English Wikipedia.}
\begin{center}
\resizebox{13cm}{!}{
\begin{tabular}{|c|c|c|}
  \hline
 Rank & PageRank persons & 2DRank persons \\
  \hline
1st & Napoleon            & Frank Sinatra         \\
2nd & Barack Obama        & Michael Jackson       \\
3rd & Carl Linnaeus       & Pope Pius XII         \\
4th & Elizabeth II        & Elton John            \\
5th & George W. Bush      & Elizabeth II          \\
6th & Jesus               & Pope John Paul II     \\
7th & Aristotle           & Beyonc\'e Knowles \\
8th & William Shakespeare & Jorge Luis Borges \\
9th & Adolf Hitler        & Mariah Carey \\
10th & Franklin D. Roosevelt    & Vladimir Putin \\
  \hline
\end{tabular}}
\end{center}
\label{tableENFigures}\label{table2}
\end{table*}

\newpage$\phantom{.}$
\begin{table*}[!ht]
\caption{List of country code (CC), countries as birth places of
historical figures, and language code (LC) for each country. LC is
determined by the most spoken language in the given country. Country
codes are based on country codes of Internet top-level domains and
language codes are based on language edition codes of Wikipedia;
WR represents all languages other than the considered 24 languages. }
\begin{center}
\resizebox{16.4cm}{!}{
\begin{tabular}{|c|c|c|c|c|c|c|c|c|}
  \hline
  CC & Country & LC & CC & Country &LC & CC & Country &LC \\
  \hline
AE & {\small United Arab Emirates} & AR & AF & {\small Afghanistan} & FA & AL & {\small Albania} & WR \\
AR & {\small Argentina }& ES  & AT & {\small Austria} & DE & AU & {\small Australia} & EN \\
AZ & {\small Azerbaijan }& TR & BE & {\small Belgium} & NL & BG & {\small Bulgaria} & WR \\
BR & {\small Brazil} &PT & BS & {\small Bahamas} & EN & BY & {\small Belarus} & RU \\
CA & {\small Canada }&EN & CH & {\small Switzerland} & DE & CL & {\small Chile} & ES\\
CN & {\small China} & ZH & CO & {\small Colombia} & ES & CU & {\small Cuba} & ES \\
CY & {\small Cyprus} & EL & CZ & {\small Czech Rep.} & WR & DE & {\small Germany} & DE\\
DK & {\small Denmark} & DA & DZ & {\small Algeria} & AR & EG & {\small Egypt} & AR\\
ES & {\small Spain} &ES & FI & {\small Finland} & WR & FR & {\small France} & FR \\
GE & {\small Georgia} & WR & GR & {\small Greece} & EL & HK & {\small Hong Kong} & ZH \\
HR & {\small Croatia} & WR & HU & {\small Hungary} & HU & ID & {\small Indonesia} & WR \\
IE & {\small Ireland} & EN & IL & {\small Israel} & HE & IN & {\small India} & HI \\
IQ & {\small Iraq} & AR & IR & {\small Iran} & FA & IS & {\small Iceland} & WR \\
IT & {\small Italy} & IT & JP & {\small Japan} & JA & KE & {\small Kenya} & EN \\
KG & {\small Kyrgyzstan} & WR & KH & {\small Cambodia} & WR & KO & {\small S. Korea} & KO \\
KP & {\small N. Korea} & KO & KW & {\small Kuwait} & AR & KZ & {\small Kazakhstan} & WR \\
LB & {\small Lebanon} & AR & LT & {\small Lithuania} & WR & LV & {\small Latvia} & WR \\
LY & {\small Libya} & AR & MK & {\small Macedonia} & WR & MM & {\small Myanmar} & WR \\
MN & {\small Mongolia} & WR & MX & {\small Mexico} & ES & MY & {\small Malaysia} & MS \\
NL & {\small Netherlands} & NL & NO & {\small Norway} & WR & NP & {\small Nepal} & WR \\
NZ & {\small New Zealand} & EN & OM & {\small Oman} & AR & PA & {\small Panama} & ES \\
PE & {\small Peru} & ES & PK & {\small Pakistan} & HI & PL & {\small Poland} & PL \\
PS & {\small State of Palestine} & AR & PT & {\small Portugal} & PT & RO & {\small Romania} & WR \\
RS & {\small Serbia} & WR & RU & {\small Russia} & RU & SA & {\small Saudi Arabia} & AR\\
SD & {\small Sudan} & AR & SE & {\small Sweden} & SV & SG & {\small Singapore} & ZH \\
SI & {\small Slovenia} & WR & SK & {\small Slovakia} & WR & SR & {\small Suriname} & NL \\
SY & {\small Syria} & AR & TH & {\small Thailand} & TH & TJ & {\small Tajikistan} & WR \\
TN & {\small Tunisia} & AR & TR & {\small Turkey} & TR & TW & {\small Taiwan} & ZH\\
TZ & {\small Tanzania} & WR & UA & {\small Ukraine} & WR & UK & {\small United Kingdom} & EN\\
US & {\small United States} & EN & UZ & {\small Uzbekistan} & WR & VE & {\small Venezuela} & ES\\
VN & {\small Vietnam} & VI & XX & {\small Unknown} & WR & YE & {\small Yemen} & AR \\
ZA & {\small South Africa} & WR &        &    &  & &    & \\
  \hline
\end{tabular}}
\end{center}
\label{tableCountry}\label{table3}
\end{table*}

\newpage$\phantom{.}$
\begin{table*}[!ht]
\caption{List of global historical figures by PageRank and 2DRank
for all 24 Wikipedia editions. All names are represented by
the corresponding article titles in the English Wikipedia. Here, $\Theta_A$ is the
ranking score of algorithm $A$ (\ref{eq1}); $N_A$ is the
number of appearances of a given person in the top 100 rank for
all editions.}
\begin{center}
\resizebox{16cm}{!}{
\begin{tabular}{|c|c|c|c|c|c|c|c|c|}
  \hline
 Rank & PageRank global figures & $\Theta_{PR}$ & $N_A$ & 2DRank
global figures & $\Theta_{2D}$ & $N_A$\\
  \hline
1st & Carl Linnaeus & 2284 & 24 & Adolf Hitler          & 1557 & 20 \\
2nd & Jesus         & 2282 & 24 & Michael Jackson       & 1315 & 17 \\
3rd & Aristotle     & 2237 & 24 & Madonna (entertainer) & 991 & 14 \\
4th & Napoleon      & 2208 & 24 & Jesus                 & 943 & 14 \\
5th & Adolf Hitler  & 2112 & 24 & Ludwig van Beethoven  & 872 & 14 \\
6th & Julius Caesar & 1952 & 23 & Wolfgang Amadeus Mozart & 853 & 11 \\
7th & Plato         & 1949  & 24 & Pope Benedict XVI & 840 & 12 \\
8th & William Shakespeare & 1861 & 24 & Alexander the Great & 789 & 11 \\
9th & Albert Eistein & 1847 & 24 & Charles Darwin & 773 & 12 \\
10th & Elizabeth II & 1789 & 24 & Barack Obama & 754 & 16 \\
  \hline
\end{tabular}}
\end{center}
\label{tableGlobalFigure10}\label{table4}
\end{table*}

\newpage$\phantom{.}$
\begin{table*}[!ht]
\caption{List of the top 10 global female historical figures by PageRank and
2DRank for all the 24 Wikipedia editions. All names are represented by
article titles in the English Wikipedia. Here, $\Theta_A$ is the
ranking score of the algorithm $A$ (Eq.\ref{eq1}); $N_A$ is the
number of appearances of a given person in the top 100 rank for
all editions. Here $CC$ is the birth country code and $LC$ is the language code of
the given historical figure.}
\begin{center}
\resizebox{12cm}{!}{
\begin{tabular}{|c|c|c|c|c|c|c|}
  \hline
 Rank & $\Theta_{PR}$ & $N_A$ &PageRank female figures & $CC$ & Century &
$LC$ \\
  \hline
1 & 1789 & 24 & Elizabeth II & UK & 20 & EN \\
2 & 1094 & 17 & Mary (mother of Jesus) & IL & -1 & HE \\
3 & 404 & 12 & Queen Victoria & UK & 19 & EN \\
4 & 234 & 6 & Elizabeth I of England & UK & 16 & EN \\
5 & 128 & 2 & Maria Theresa & AT & 18 & DE \\
6 & 100 & 1 & Benazir Bhutto & PK & 20 & HI \\
7 & 94 & 1 & Catherine the Great & PL & 18 & PL \\
8 & 91 & 1 & Anne Frank & DE & 20 & DE \\
9 & 87 & 1 & Indira Gandhi & IN & 20 & HI \\
10 & 86 & 1 & Margrethe II of Denmark & DK & 20 & DA \\
  \hline
Rank & $\Theta_{2D}$ & $N_A$ &2DRank female figures & $CC$ & Century & $LC$
\\
\hline
1 & 991 & 14 & Madonna (entertainer) & US & 20 & EN \\
2 & 664 & 9 & Elizabeth II & UK & 20 & EN \\
3 & 580 & 8 & Mary (mother of Jesus) & IL & -1 & HE \\
4 & 550 & 9 & Queen Victoria & UK & 19 & EN \\
5 & 225 & 5 & Agatha Christie & UK & 19 & EN \\
6 & 211 & 4 & Mariah Carey & US & 20 & EN \\
7 & 206 & 7 & Britney Spears & US & 20 & EN \\
8 & 200 & 3 & Margaret Thatcher & UK & 20 & EN \\
9 & 191 & 2 & Martina Navratilova & CZ & 20 & WR \\
10 & 175 & 2 & Elizabeth I of England & UK & 16 & EN \\
\hline
\end{tabular}}
\end{center}
\label{tableWomen}\label{table5}
\end{table*}

\newpage$\phantom{.}$
\begin{table*}[!ht]
\caption{Numbers of certain
historical figures for top 100 list of each language:
$N_1$ is the number of historical figures of a given
language among the top 100 PageRank global historical figures;
$N_2$ is the number of historical figures of a given
language among the top 100
PageRank historical figures for the given language edition;
$N_3$ is the number of historical figures of a
given  language among the top 100 2DRank
global historical figures;
$N_4$ is the number of historical figures of
a given language among the top 100 2DRank historical figures for the given
language edition.}
\begin{center}
\resizebox{12.0cm}{!}{
\begin{tabular}{|c|c|c|c|c|c|c|c|c|c|}
  \hline
  Language & $N_1$ & $N_2$ & $N_3$ & $N_4$ & Language & $N_1$ & $N_2$ &
$N_3$ & $N_4$  \\
  \hline
EN & 22 & 47 & 27 & 64 & RU & 2 & 29 & 3 & 27 \\
NL & 2 & 10 & 4 & 38 & HE & 2 & 17 & 2 & 22 \\
DE & 20 & 41 & 16 & 55 & TR & 2 & 27 & 2 & 54 \\
FR & 8 & 33 & 3 & 32 & AR & 8 & 42 & 5 & 69 \\
ES & 2 & 20 & 5 & 39 & FA & 0 & 46 & 1 & 64 \\
IT & 11 & 31 & 9 & 43 & HI & 1 & 65 & 0 & 76 \\
PT & 0 & 19 & 0 & 35 & MS & 0 & 15 & 0 & 40 \\
EL & 5 & 28 & 2 & 55 & TH & 0 & 46 & 0 & 53 \\
DA & 0 & 31 & 1 & 48 & VI & 0 & 7 & 0 & 30 \\
SV & 1 & 26 & 1 & 39 & ZH & 5 & 43 & 6 & 79 \\
PL & 1 & 20 & 2 & 26 & KO & 0 & 34 & 0 & 59 \\
HU & 0 & 18 & 0 & 18 & JA & 0 & 41 & 4 & 80 \\
WR & 8 & - & 7 & - & & & & & \\
\hline
\end{tabular}}
\end{center}
\label{tablelocalfigures}\label{table6}
\end{table*}

\end{document}